\documentstyle[aps,prb,epsfig,multicol,float]{revtex}

\begin{document}

\title{Mixed moment wave function for magnetic heavy fermion compounds}

\author{Yunkyu Bang$^{*}$}
\address{
Theoretical Division Los Alamos National Laboratory, Los Alamos, New Mexico \
87545 }


\maketitle

\begin{abstract}
We propose a variational wave function for the ground state of the magnetic
heavy fermion (HF) systems, in which both the Kondo and the RKKY interactions
are variationally incorporated and the local f-orbital state exists as a linear
combination of a full local moment state and a fully compensated state (mixed
moment state). We describe the  mechanism for the mixed moment ground state
based on the large-N treatment of the Kondo lattice Hamiltonian added with RKKY
interaction. With the mixed moment ground state  we can explain several
puzzling experiments in magnetic HF compounds such as a small value of local
moment, coexistence of the antiferromagnetic (AFM) and the paramagnetic (PM)
phases, local quantum criticality, etc.
\end{abstract}

\noindent

\begin{multicols}{2}

\section{Introduction}
As a conventional wisdom, one often employs the Kondo mechanism to understand
the HF behavior, in which the conduction electrons make bound states with the
local f-orbital moments in a coherent fashion to result in fermionic
quasiparticles integrating the local f-orbital moments with it. While it is
still  unclear how the single site Kondo mechanism can be generalized to the
periodic lattice system, at least the above line of thinking seems to provide
the correct energy scale, i.e. the coherent energy scale of forming heavy
renormalized fermionic quasiparticles \cite {Read_Tk}. However, this
phenomenological picture doesn't explain many experimental observations of HF
systems, such as, strong temperature dependence of specific heat $C(T)$ and
bulk susceptibility $\chi(T)$ below the coherent temperature $T_{coh}$, the
coexistence of the magnetism (often antiferromagnetism (AFM)) and the heavy
quasiparticle in some compounds, and superconductivity in some other compounds.
Those compounds which displays both HF properties as well as the magnetism
(either magnetic fluctuations or the magnetic long range order(LRO)) are
conveniently called "magnetic HF" by experimentalists. Then again the common
wisdom for this class of HF systems is the competition between the Kondo
coupling between the local f-orbitals and the conduction electrons and the RKKY
interaction among the local f-orbitals themselves \cite{Doniach_1977}. And at
qualitative level the result of this competition is that the ground state
should be either pure HF state (all local moment is completely compensated by
conduction electrons) or pure magnetic state (no compensation of the local
moment) depending on which energy scale is larger.
%
However the experimental situation is that these magnetic HF compounds not only
show the coexistence of the local moment and the heavy quasiparticle but also
almost always the size of the local moment is much smaller than its ionic limit
value.
Therefore, it is clear  that  a minimum requirement to understand experimental
observations is that the true ground state wave function for the magnetic HF
compounds should simultaneously possess both the heavy quasiparticle and local
moment characters.

In this paper, we propose a variational wave function for the ground state of
the magnetic HF systems, in which the local f-orbital wave function exists in
the mixed states (a linear combination of the full local moment state and the
fully compensated state), and we call it "mixed moment state" after the "mixed
valence state"\cite{MV}. In the following sections, we show how this mixed
moment state can be realized in the Kondo lattice model using the large-N
scheme. This part is rather qualitative based on account of the energetics of
the competition between  different couplings. Once this mixed moment state is
phenomenologically accepted, then this ground state properties can immediately
provide explanations for the most puzzling questions of magnetic HF compounds:
small uncompensated local moment, the coexistence of the AFM  and  PM domains
(the PM domain can be even a superconducting phase), and as the most
interesting possibility, the local quantum criticality coinciding with the
magnetic criticality and its relation with the Fermi surface (FS) fluctuations,
etc.

\section{the mixed moment state}

We start with the Kondo lattice Hamiltonian (KLH).

\begin{eqnarray}
H & = & \sum_{k,m} \epsilon_{k,m} c\dag _{k,m} c_{k,m} + \epsilon^0 _f
\sum_{i,m} f\dag _{i,m} f_{i,m} \\ \nonumber & & + \frac{J_{Kondo}}{N} \sum_{i}
\vec{S}_{i} \cdot \vec{\sigma} _{i} + \frac{I_{RKKY}}{N^2} \sum_{<ij>}
\vec{S}_{i} \cdot \vec{S}_{j}
\end{eqnarray}

\noindent where $c\dag _{k,m}, c_{k,m}$ is the conduction electron creation and
annihilation operators and $f\dag_{i,m}, f_{i,m}$ are the local f-orbital
operators, and the spin operators are the large N generalization of SU(2) spin
$\vec{S}_i=f\dag _{i,m} \vec{\tau} f_{i,m^{'}}$ and $\vec{\sigma}_i =c\dag
_{i,m} \vec{\tau} c_{i,m^{'}}$ for the f-orbital  and conduction electron
spins, respectively. The above Hamiltonian is the usual Kondo lattice
Hamiltonian extended with the large N spin degeneracy ($m=1,...,N$) and the
f-orbital occupancy is constrained with $n^{f} _{i}=\sum_{m} f\dag _{i,m}
f_{i,m}=N q_0$ \cite{Nq0} representing the large on-site Coulomb interaction
between the localized  f-orbitals at the same site. Finally the RKKY
interaction term between the f-orbitals is added \cite{comment_RKKY}. The above
Hamiltonian has been studied by numerous authors using various
techniques\cite{Auerbach}. Without the RKKY term, the various approximation
(mainly the approximation treating the constraint $n^{f} _{i}=N q_0$) provides
a solution of heavy renormalized coherent band(s) to explain the HF phenomena.
When the RKKY term is added as in Eq.(1) not only any approximate solution
becomes more complicated but more importantly the question of the correctness
of treating the constraint  becomes crucial because the fate of the
competition/interplay between the Kondo and the RKKY interactions is largely
unknown to determine the true ground state.
There are a few studies of the two impurity version of this model
\cite{two_imp}. All these studies indicate that there is a critical ratio of
interactions ($J_{Kondo}/I_{RKKY}$) from which the ground state flows away
toward either a magnetically correlated state or the  Kondo singlet state.
While we learn from these studies that there is an interesting competition
between the Kondo and RKKY interactions, a naive extension of this picture to
the lattice system is not warranted.

As discussed in the introduction, motivated by experimental observations, we
construct a new  variational wave function for the ground state of the above
Hamiltonian. We start with the f-orbital wave function as follows.

\begin{equation}
|f_{i,m}>= \sqrt{1-\alpha} |f_{i,m}>_{iti} + \sqrt{\alpha} |f_{i,m}>_{loc}
\end{equation}

 The above expression is designed to indicate that the  f-orbital state in the
ground state can be a superposition of two qualitatively different states; one
is the itinerant state ($|f_{i,m}>_{iti}$) which makes a coherent singlet state
with the conduction electrons via renormalized hybridization through Kondo
coupling \cite{Auerbach} and the other is the full local moment state
($|f_{i,m}>_{loc}$) which remains intact from forming Kondo singlet or heavy
quasiparticle but couples only as a local magnetic  moment with other
electrons.
 With the above Ansatz for the f-orbital state, we diagonalize the Hamiltonian Eq.(1)
in large N approximation. The  ground state should determine the value of
$\alpha$ to minimized the total energy of the Hamiltonian. This wave function,
if it is realized as a true ground state, displays the mixed moment state in a
basically same manner as the mixed valence state \cite{MV}, and the origin of
these mixed states is the local Coulomb interaction.

Assuming the variational Ansatz of Eq.(2), the Kondo coupling term in Eq.(1) is
decomposed into two parts depending on which component of f-orbital states is
involved.

\begin{eqnarray}
\frac{J_{Kondo}}{N} \sum_{i} \vec{S}_{i} \cdot \vec{\sigma} _{i}  = (1-\alpha)
\frac{J_{Kondo}}{N} \nonumber
\\  \cdot \sum_{i,m,m^{'}} f\dag _{i,m} c_{i,m} c\dag
_{i,m^{'}} f_{i,m^{'}} + \alpha \frac{J_{Kondo}}{N} \sum_{i} \vec{S}_{i} \cdot
\vec{\sigma} _{i}
\end{eqnarray}

 In the above equation, we rewrite the
Kondo term coupled with the itinerant component of the  f-orbital by the four
fermionic expression often employed in the large-N treatment of Kondo coupling
\cite{Auerbach}. The second term represents the local moment part of the
f-orbital state, which remains as a local moment and does not participate in
forming the heavy quasiparticles.
Now the Hamiltonian can be written as,

\begin{eqnarray}
H & = & H_{Kondo} + H_{mag}, \\
H_{Kondo} & = & \sum_{k,m} \epsilon_{k,m} c\dag _{k,m} c_{k,m} + \epsilon^0 _f
\sum_{i,m} f\dag _{i,m} f_{i,m} \\ \nonumber & & + (1-\alpha)
\frac{J_{Kondo}}{N} \sum_{i,m,m^{'}} f\dag _{i,m} c_{i,m} c\dag _{i,m^{'}}
f_{i,m^{'}}, \\
H_{mag} & = & \alpha \frac{J_{Kondo}}{N} \sum_{i} \vec{S}_{i} \cdot
\vec{\sigma} _{i}+  \alpha^2 \frac{I_{RKKY}}{N^2} \sum_{<ij>} \vec{S}_{i} \cdot
\vec{S}_{j}
\end{eqnarray}

$H_{Kondo}$ part is well studied by many authors and in particular the large N
technique is convenient to produce the heavy renormalized quasiparticle
band(s). Switching on the Kondo coupling ($J_{Kondo}$), the ground state of
$H_{Kondo}$ lowers its energy compared to $J_{Kondo}=0$ state ($\Delta
E_{Kondo}$). $H_{mag}$ part also should gain energy by developing a magnetic
correlation, but the first term of $H_{mag}$, by coupling with the conduction
electrons ($c$ electrons), also increases the kinetic energy of the
$H_{Kondo}$.
 First, we consider $H_{Kondo}$. The only difference from the
previous studies is that the Kondo coupling $J_{Kondo}$ is replaced by
$(1-\alpha) J_{Kondo}$ and also  the f-orbital occupancy constraint is modified
as $n^{f} _{i}= (1-\alpha) Nq_0$ (accordingly the Fermi surface volume changes
with $\alpha$). Introducing the Stratonovich-Hubbard decoupling field
$\phi_0=(1-\alpha) \frac{J_{Kondo}}{N} \sum_{m} <f\dag _{i,m} c_{i,m}>$,
$H_{Kondo}$ is diagonalized and the renormalized $\epsilon_{f}$ is fixed to
satisfy the constraint $n^{f} _{i}= (1-\alpha) Nq_0$. Following Read et al.
(Ref.[2]), the Kondo energy gain $\Delta E_{Kondo}$ is written as,

\begin{equation}
\Delta E_{Kondo} = N \rho_0 \phi_0 ^2  \ln [\epsilon_f/D] -\epsilon_f (1-\alpha) +
\frac{N }{ J_{K} (1-\alpha)} \phi_0 ^2
\end{equation}

\noindent where $D$ is the typical conduction band width and $\rho_0$ is the
density of states (DOS) of it. The self-consistent equations are as follows.

\begin{eqnarray}
N \rho_0 \phi_0 ^2 \frac{1}{\epsilon_f} &=& (1-\alpha) \\
\epsilon_f &=& D \exp [\frac{-1}{\rho_0 J_{K} (1-\alpha)}].
\end{eqnarray}
Using the above equations, we find the Kondo energy gain per site  is the
following.

\begin{equation}
\Delta E_{Kondo} = - (1-\alpha) D \exp [\frac{-1}{\rho_0 J_{K} (1-\alpha)}].
\end{equation}

The system gains the above energy by compensating a fraction  $(1-\alpha)$ of
the local moments  via Kondo screening and the conduction band becomes a
renormalized heavy band(s). Now let us consider $H_{mag}$. The RKKY term would
gain the following energy per site with the maximum polarization $<\vec{S} _i>=
S_z =\frac{N-1}{2}$.

\begin{equation}
\Delta E_{RKKY} = - \alpha^2 \frac{I_{RKKY}}{N^2} S^2 _z \sim O(N^0).
\end{equation}

The first term of $H_{mag}$ (Eq.(6)) now forces the AFM coupling of the
conduction electrons ($c$ electrons) with a staggered field $h^z _i$ or a
staggered energy level $\Delta_i=\alpha \frac{J_{K}}{N} S_z$. With this
staggered energy level for $c$-electrons, the renormalized heavy band(s) should
further develop spin density wave (SDW) ordering and it will cost the kinetic
energy increase as follows.

\begin{equation}
\Delta E_{kin} = \frac{1}{2} \rho_{HF}~ u^2 _{k_F} \Delta_i ^2 +
\frac{\rho_{HF}}{v^2 _F (2 k_F -Q)^2} \Delta_i ^4  \ldots.
\end{equation}

\noindent where $\rho_{HF}$ is the DOS of the renormalized band at chemical
potential and $u_{k}$ is the Bogoliubov coefficient of $c_{k}$ component of the
renormalized band operators which diagonalize Eq.(5). $v_F$ is the Fermi
velocity and $Q$ is the SDW ordering vector \cite{Kinetic}.
Finally, the total energy gain of $H_{mag}$ with both the local magnetic and
SDW orderings is as follows.

\begin{equation}
\Delta E_{mag}=-\rho_{HF}~ u^2 _{k_F} \Delta_i ^2 - \alpha^2
\frac{I_{RKKY}}{N^2} S^2 _z
\end{equation}

Now collecting all the energy gain and loss, the total energy difference by the
Kondo and RKKY couplings  is written as

\begin{eqnarray}
\Delta E_{total} &=& \Delta E_{Kondo}+\Delta E_{kin}+\Delta E_{mag} \\
&= & -(1-\alpha) D \exp [\frac{-1}{\rho_0 J_{Kondo} (1-\alpha)}] \nonumber \\ &
& + A^{(2)} \alpha^2 + A^{(4)}  \alpha^4 + ...
\end{eqnarray}
where $A^{(2)}= -\frac{1}{2} \rho_{HF}~ u^2 _{k_F} (\frac{J_{Kondo}}{N})^2 S^2
_z - \frac{I_{RKKY}}{N^2} S^2 _z$, and $A^{(4)}= \frac{\rho_{HF}}{v^2 _F (k_F
-Q)^2} (\frac{J_{Kondo}}{N})^4 S^4 _z$. Note that $A^{(2)} <0 $, $A^{(4)}  >0 $
and since $S_z \sim$ O(N) they are all O(N$^0$). Ellipsis indicates the higher
order terms (O($\alpha^6$) etc.)

Treating $A^{(2)} $ and $A^{(4)}$ as phenomenological parameters, in Fig.(1) we
show the schematic total energy gain $\Delta E_{total} (\alpha)$ of Eq.(15) for
the three representative cases. There is always a local minimum with a finite
value of $\alpha$ (due to $A^{(2)} <0 $, $A^{(4)}  >0 $ ). However, when this
local minimum is not a global minimum as in Fig.1.a, the ground state is a pure
HF state ($\alpha=0$, no magnetic HF) with  fully screened local f-moments. In
contrast, when this local minimum with a finite $\alpha$ becomes a global
minimum as depicted in Fig.1.b., then the ground state is a "mixed moment"
state. If a fine tuning occurs such as using pressure, magnetic fields,
chemical substitution, etc., then two minima can be degenerate with $\alpha\neq
0$ and $\alpha=0$ as in Fig.1.c. In this case, as the tuning parameter changes,
the ground state goes through a first order transition accompanying with a jump
of magnetization and FS volume. Finally, the above discussions are mean field
level and therefore a $\alpha$  smaller than a critical value would not develop
a true long range order (LRO) of local moments due to fluctuations; it means
that the nonmagnetic phase near the magnetic phase boundary should have a small
but finite $\alpha$ value.

\section{Neutron diffraction (ND) and NMR  experiments in URu$_2$Si$_2$}
In our mixed moment state,  the ground state is a superposition of the
unscreened full local moments and the Kondo renormalized itinerant band(s). We
first consider the consequence of this ground state for the measurement of the
local moment by neutron diffraction (ND) experiment. At any given time, a
fraction ($\alpha$) of U sites have a full local moment and a fraction
($1-\alpha$) of U sites are completely quenched by the conduction electrons.
Now the unquenched local moment is magnetically screened by the opposite SDW
moment developed by the renormalized itinerant band. This incompletely screened 
magnetic moment is the effective
local moment ($S_{z} ^{eff}=S_{z}-m_{z}$, where
$m_z=n_{cond}(\uparrow)-n_{cond}(\downarrow)$). Now the ND should see the
ensemble averaged size of the local moment, i.e. $\mu_{eff}=g \mu_B S_{z}
^{eff} \times \alpha$. This value can be tiny ($\mu_{eff} \sim 0.03 \mu_B$ for
URu$_2$Si$_2$) while $S_{z} ^{eff}$ and $m_z$ are not necessarily tiny.

A recent NMR  experiment by K. Matsuda et al. \cite{Matsuda} reports another
puzzling data on magnetic moment. In this experiment, $^{29}$Si NMR spectra
clearly shows that $^{29}$Si sites see both the AFM ordered phase and the PM phase
below the magnetic transition temperature $T_0$ and these authors interpret the
data with a spatial inhomogeneous mixture of AFM and PM domains. From the
estimated volume fraction of AFM domain, it is concluded that the actual size
of local moment is at least $0.3 \mu_B$/U, an order of magnitude larger than
the estimation from the neutron Bragg peak intensity\cite{ND}. In our picture,
the coexistence of AFM and PM is not of a static inhomogeneous domain
structure, but at any given time the $^{29}$Si nuclei should see a fraction
($\alpha$) of $S_{z} ^{eff}$ and a fraction ($1-\alpha$) of zero moment from U
sites. From this experiment we can read the size of SDW moment by $S_{z}
^{eff}=S_{z}-m_{z}= 0.3 /g$ and $ S_{z}=\mu_{para}/ \mu_B=1.2/g$. Combining ND
and NMR data, $\mu_{eff} \sim 0.03 \mu_B$ = ($0.3 \mu_B \times \alpha$),
$\alpha \sim 0.1$ is estimated  for URu$_2$Si$_2$ for ambient pressure.
Also the observation of a distribution of local effective fields at $^{29}$Si
sites from $^{29}$Si NMR by O.O. Bernal et al. \cite{Bernal} is a natural
consequence of the local moment fluctuations of nearby U sites between zero and
a finite value.

A strong pressure dependence of the local moment size from ND by H. Amitsuka et
al. \cite{Amitsuka} (from $0.017 \mu_B$ to $0.25 \mu_B$) and much slower
increase of the magnetic ordering temperature ($T_m$) is not inconsistent with
our model:  the measured local moment size is (U full local moment $-$  SDW
moment) $\times \alpha$  and in zeroth approximation $T_m$ is $\sim \alpha^2$
(Eq.(6))  assuming the RKKY coupling $I_{RKKY}$ constant \cite{I_RKKY}.
Applying pressure should change $\alpha$ (increasing in URu$_2$Si$_2$) and
$T_{m}$ increases, but the increase of the effective local moment size should
be a more complicate function of $\alpha$.

\section{Fermi surface volume and $C(T)$ jump}
In the mixed moment state, the total carrier density in the renormalized
itinerant band is given as $n_{tot}=n_{c}+Nq_0 (1-\alpha)$ ($n_c$ is the
original conduction electron density). Therefore, the FS volume changes by how
much of the f-electrons participates to form the coherent heavy quasiparticle
band(s) through the Kondo coupling. As a result the fluctuations of the local
moment weight $\alpha$ is intimately coupled to the fluctuations of FS and any
phase transition of one part should trigger a transition of the other part. One
consequence is an enhanced specific heat jump $\Delta C(T)$ at the second order
transition as commonly observed in many magnetic HF compounds. In general, the
specific heat jump at the second order transition temperature can be expressed
as follows.

\begin{eqnarray}
\Delta C_{mag}/T  =  - \frac{\partial ^2 F}{\partial T^2}  & \sim &
a_1 \frac{\partial  m_{loc} ^2}{\partial  T} + a_2 \frac{\partial
\Delta^2_{SDW}}{\partial T}  \\
\Delta C_{SC}/T   & \sim & b_1 \frac{\partial \Delta^2 _{sc}}{\partial T} + b_2
\frac{\partial \alpha^2}{\partial T}
\end{eqnarray}

The first case ($m_{loc}= \alpha S^z$) can be applied to URu$_2$Si$_2$, which
is well known for its extremely tiny local moment ($\mu_{eff} \sim 0.03 \mu_B$)
from ND  but displaying a huge jump ($\Delta C/T_0 \sim 300 mJ/K^2$ mol) at the
magnetic transition $T_0 =17.5K.$ To obtain some quantitative scales, first
note that $T_0=17.5K$ indicates the RKKY coupling should  be large enough to
justify the relatively high transition temperature with a small averaged local
moment value ($\alpha S^z$), and it also means that $J_{Kondo}$ is not a small
value \cite {I_RKKY} and indeed the experimental $T_K \sim 60 K$.
 Thus  the average
internal fields induced by the local moment ordering is given by $h^z _i=
\frac{\alpha}{g \mu_B} \frac{J_{Kondo}}{N} S_z $ (the first term in Eq.(6)),
which is not so tiny value despite the acclaimed small effective local moment.
Then the itinerant renormalized band develops a SDW ordering over a part of FS
simultaneously at $T=T_0$ and it can provide the major contribution to the
specific heat jump (in the mean field theory $a_2 \sim \rho_{HF}$).
The second case (Eq.(17)) is for the SC transition (for example,
CeCoIn$_5$\cite{Lengyel}). The deformation of FS due to the SC should trigger a
concomitant change in $\alpha$.

\section{Coincidence of local quantum criticality and the magnetic
criticality}
Many magnetic HF compounds exhibit non-Fermi liquid (NFL) behaviors in
resistivity, specific heat, neutron scattering, susceptibility etc
\cite{Stewart}.  Often these behaviors coincide with the  magnetic quantum
transition (mostly AFM transition). However, theoretically any quantum critical
fluctuations, which is spatially correlated, is not sufficient to explain the
NFL behaviors. Therefore, what is naturally required is a "local quantum
criticality" not only for NFL HF compounds but perhaps also for the high
temperature superconducting materials.

There are a few theoretical proposals\cite{Si,Varma} of local quantum
criticality in the strongly correlated electron system with different ideas.
Our "mixed moment" ground state can provide a natural mechanism for the local
quantum criticality, in particular, for NFL HF compounds. The first term of
Eq.(6) provides a coupling between the conduction electrons ($\vec{\sigma_i}$)
and local spins ($\vec{S_i}$), and the local spins develop its own dynamics
through the RKKY interaction, which can go through the magnetic quantum
criticality by tuning external parameters. This is a two component spin-fermion
model recently proposed to study the superconductivity of CeMIn$_5$\cite{Bang}.
Now in addition to the spatially correlated magnetic fluctuations, if the
moment weight fluctuations $<\delta \alpha_i  \delta \alpha_j>$ is allowed in
higher order corrections, this fluctuations absorbs any value of momentum
exchange just like impurities as far as the typical energy of this fluctuations
is much higher than the typical low energy scale of the spin fluctuations $<S_i
S_j>$.
Then all conduction electrons around FS can be scattered off each other by a
critical magnetic fluctuations with a help of the local moment weight
fluctuations. This process is depicted diagramatically in Fig.2. Therefore, in
this picture {\em the spatially correlated magnetic quantum criticality becomes
a local quantum criticality with a  help of the local moment weight
fluctuations.}

\section{Conclusion}

To summarize, we propose a new variational  wave function for the magnetic HF
which represents a mixed moment ground state. We use a mean field analysis in
large N limit of  the KLH added with RKKY interaction and show that this is
indeed a generic ground state for a certain ratio of the couplings of Kondo and
RKKY. Then we show that this mixed moment ground state can immediately provide
natural explanations for the most puzzling observations in magnetic HF
compounds ( URu$_2$Si$_2$, UPt$_3$, CeMIn$_5$, CeCu$_{6-x}$Au$_{x}$, etc.) such
as a tiny ordered magnetic moment, large specific heat jump, a coexistence of
AFM and PM phases, a distribution of internal fields, and most interestingly
the local quantum criticality coinciding with the spatial magnetic criticality.
More quantitative calculations for each of these properties will be reported in
separate publication.

We acknowledge our thanks to A.V. Balatsky, W. Bao and  J.D. Thompson for useful
discussions. This work was supported by US DoE and partially by the Korean
Science and Engineering Foundation (KOSEF) through the Grant No.
1999-2-114-005-5.

\begin{figure}
\epsfig{figure=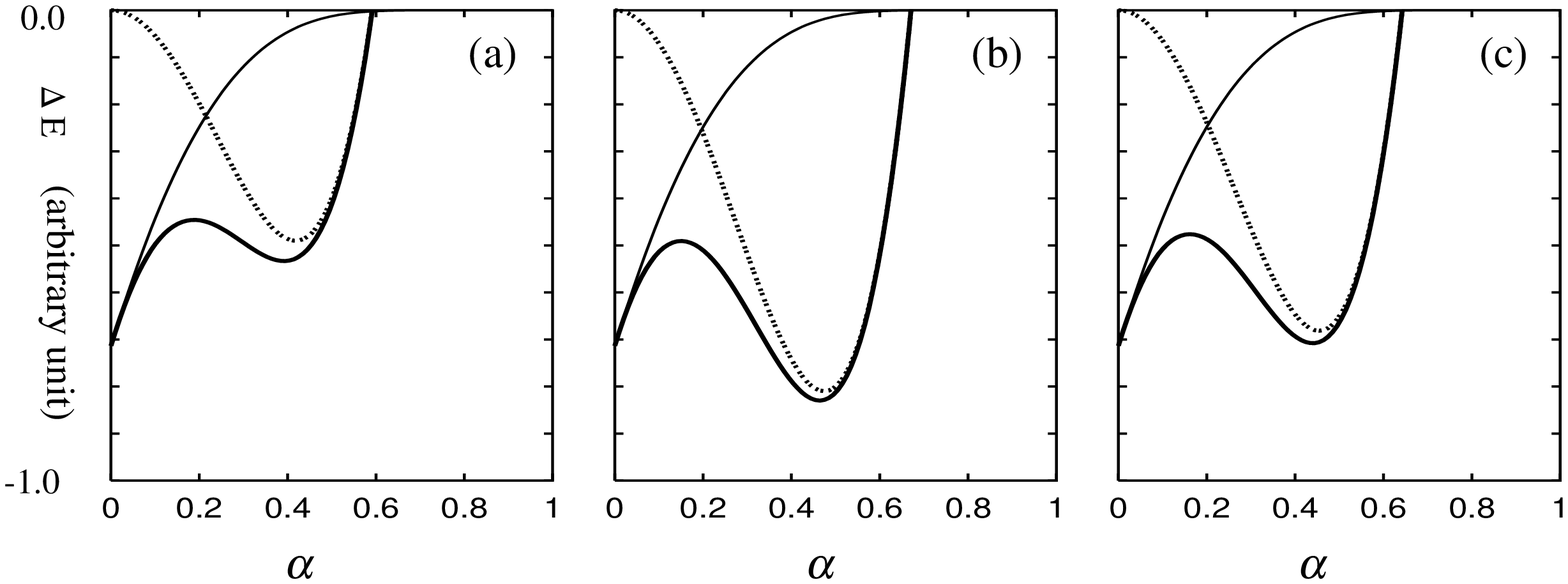,width=1.0 \linewidth} \caption{Energy gains as a
function of $\alpha$ for the representative cases of different ratios of Kondo
and RKKY couplings: $\Delta E_{tot}(\alpha)$ (solid line), $\Delta
E_{Kondo}(\alpha)$ (thin solid line), $\Delta E_{mag}(\alpha)+\Delta
E_{kin}(\alpha)$ (dotted line). \label{fig1}}
\end{figure}

\begin{figure}
\epsfig{figure=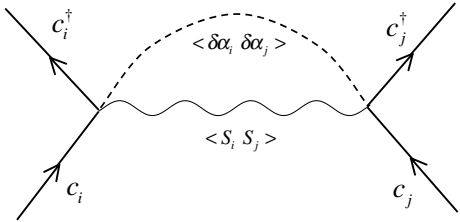,width=1.0\linewidth} \caption{Feynman diagram for
the vertex of electron electron interaction mediated by magnetic fluctuations.
\label{fig2}}
\end{figure}

\end{multicols}


\begin{references}
\bibitem{Nq0}
As usual, $q_0$ is treated as an $O(N^0)$ constant, but in the end $Nq_0$ sets
equal to 1. This technical detail doesn't matter for us.
\bibitem{comment_RKKY}
The RKKY term is written as one order  higher in 1/N ($\sim O(N^0)$) than the
other terms ($\sim O(N^1)$). This is necessary later for the meaningful limits
of the kinetic  and  magnetic energies in competition.
\bibitem{Kinetic}
$\Delta E_{kin}$ is estimated based on the SDW ordering by a staggered energy
level $\Delta_i$ on the itinerant band. But developing SDW would reduce some
part of the Kondo screening, which then would increases the energy. While all
this feedback effect is neglected here, based on a general stability condition,
the quadratic and quartic coefficients in $\Delta E_{kin}$ are positive.
\bibitem{I_RKKY}
The RKKY coupling between local moments is estimated as $I_{RKKY} \sim
J_{Kondo} ^2 \cdot \rho_{HF} \cdot u^4 _{k_F} \cdot F(r)$; $F(r)$ is the range
function of the RKKY interaction, which is basically the  real space Fourier
transform of the polarization function of the renormalized heavy band
$\chi(q,\omega=0)$.
\end{references}
\end{document}